# Realization of the Switching Mechanism in Resistance Random Access Memory (RRAM™) Devices: Structural and Electronic Properties Affecting Electron Conductivity in Halfnium Oxide-Electrode System through First Principles Calculations

Susan Meñez Aspera[1], Hideaki Kasai[1], Hirofumi Kishi[1], Nobuyoshi Awaya[2], Shigeo Ohnishi[2], and Yukio Tamai[2]

[1]Department of Applied Physics, Osaka University, Suita, Osaka 565-0871, Japan

[2]Corporate Research and Development Group, Sharp Corporation, 1 Asahi, Daimon-cho, Fukuyama 721-8522, Japan

Email: kasai@dyn.ap.eng.osaka-u.ac.jp

Resistance Random Access Memory (RRAM™) device, with its electrically induced nanoscale resistive switching capacity, has been gaining considerable attention as future non-volatile memory device. Here, we propose a mechanism of switching based on oxygen vacancy migration-driven change in electronic properties of the transition metal oxide (TMO) film stimulated by set pulse voltages. We used density functional theory (DFT)-based calculations to account for the effect of oxygen vacancy and its migration on the electronic properties of $HfO_2$ and $Ta/HfO_2$ systems, and thereby create the entire story on RRAM™'s switching mechanism. Computational results on the activation energy barrier for oxygen vacancy migration were found to be consistent with the results of set and reset pulse voltage obtained from experiment.

Understanding of this mechanism would be beneficial to effectively realize materials design in these devices.

**INTRODUCTION**

Technological advancement in materials design is apparently geared towards the development of materials for devices that are miniature, perform intended function faster, productively cost efficient, and can be operated with low power. In the development of non-volatile memory, metal-insulator-metal device, usually termed as Resistance Random Access Memory ($RRAM^{TM}$)[1-8] device, is seen to be very promising due to its scalability and ability to be operated with low power. Materials such as transition metal oxides (TMOs) used as insulator sandwiched between two metallic electrodes can switch properties between an insulating (a high resistance state or the OFF state) and a metallic material (a low resistance state or the ON state)[9-11].

Related studies to $RRAM^{TM}$'s switching mechanism suggest that the ON and OFF mechanism is related to the creation of a conduction path on the insulator connecting one electrode to the other by either cation migration or anion migration[7]. Switching mechanism through anion migration suggests that the creation of a conduction path through defect formation and charge carrier trapping might be responsible for the switching[12-16] but has not yet come up with conclusive clarifications. Theoretical understanding behind this mechanism is of great interest to our research group whereby investigations were also conducted on the oxidative reaction of the anode through experimental and theoretical approaches[17-21].

In this study, we are interested in the conduction path formation through anion migration and the corresponding change in the electronic properties of the TMO films arising from the presence of these conduction paths. We investigate the corresponding change on the electronic properties of the TMO $HfO_2$ resulting from oxygen vacancies and charge carrier trapping around the oxygen vacancies for clarification of the switching mechanism. We will also propose a switching mechanism based on the resulting variation in the electronic properties from the given conditions.

**COMPUTATIONAL METHOD**

All calculations were performed within the Density Functional Theory (DFT) framework using the spin-polarized version of the Vienna *ab initio* simulation package (VASP)[22-25] with periodic boundary conditions in three directions. Nonlocal correction in the form of the generalized gradient approximation (GGA)[26] was included for the exchange-correlation functional. The calculations used the projector augmented wave (PAW)[27,28] potential to describe the electron-ion interaction with a cutoff energy of 400eV. A 1x1x15 k-point mesh generated by the Monkhorst-Pack scheme was used for the bulk $HfO_2$[29], while a 5x5x1 k-point mesh for the electrode/$HfO_2$. The representation of the supercell used to model the bulk $HfO_2$ systems without and with oxygen vacancy row is depicted in Figures 1a and 1b, while the system with Ta electrode in contact with bulk $HfO_2$ is depicted in Figure 4. Bulk $HfO_2$ is modeled by a supercell containing 96 atoms. We investigate the structure with oxygen vacancy row by removing two oxygen atoms, which corresponds to defect concentrations of 3.13%, from a supercell, as indicated in Fig. 1b. The electrode/$HfO_2$ slab is modeled by a supercell containing 28 atoms. Effects of electron correlations beyond GGA were taken into account within the framework of GGA+U and the simplified (rotationally invariant) approach by Dudarev et al.[30]. The Coulomb repulsion $U = 7eV$ and the local exchange interaction $J = 1eV$ were applied to describe the on-site interactions in the Hf[31]. The GGA+U calculations of electronic density of states of $HfO_2$ lead to the energy gap of 5.8eV which is consistent with a result obtained from experiment[32].

**RESULTS AND DISCUSSIONS**

**Electronic Properties and the Switching Mechanism**

The switching mechanism of RRAM[TM] basically relies on the transition between the insulator to metal property of the TMO films sandwiched by the two metallic electrodes in RRAM[TM] devices. In this paper, we presume that this switching mechanism is attributed to the formation of an oxygen vacancy row and the occurrence of charge carrier trapping along the oxygen vacancy row of the TMO films. We therefore investigate the properties of bulk $HfO_2$ without oxygen vacancy, with an oxygen vacancy row and with an oxygen vacancy row and charge carrier trapping as shown in figures 2a, 2b and 2c. Density of states analyses on this systems show that the bulk $HfO_2$ has insulator-like properties through the appearance of a large band gap around the Fermi level. On the other hand, the resulting density of states for the bulk $HfO_2$ with the oxygen vacancy rows show shifts of the energy level and the occurrence of small states near the Fermi energy in the system. Finally, by adding an extra electron per unit cell as the reference of the charge carrier trapping, the resulting density of states confirms the appearance of states around the Fermi energy for the bulk $HfO_2$ systems. These analyses indicate a change of the electronic properties of $HfO_2$ from having an insulator-like to metallic-like properties due to the presence of an oxygen vacancy row with charge carrier trapping. Moreover, we confirmed these electronic changes through the band structures of the bulk $HfO_2$ with the oxygen vacancy row and charge carrier trapping, as shown in Fig. 2d. The appearance of a band crossing the Fermi level[33] confirms electrical conductivity hence the metallic property of the system. Electron density distribution (Fig. 2e) along this band further shows that concentration of these conducting states is within the vicinity of the vacancies. Somehow, the presence of these vacancies creates trapping regions to which the excess electrons may reside and eventually acts as pathways for electron transport thus the conductivity. Figure 3 shows the band structure of a bulk $HfO_2$ with discontinuity within the rowed oxygen vacancy through the intrusion of an oxygen atom within the row of vacancies. This shows that discontinuity within the aforementioned oxygen vacancy row changes the properties of the bulk $HfO_2$ back to insulator as confirmed by the disappearance of the band that overlaps with the Fermi level. Therefore, presence of oxygen vacancy row with charge carrier trapping creates a transition from insulator to metal and the removal of the rowed oxygen vacancy thru insertion of an oxygen atom along the row converts the system back to insulator.

The effect of electrodes on these transition properties was also investigated through the $Ta/HfO_2$ electrode-bulk interface systems. Some studies related to $RRAM^{TM}$ suggest that the transition from insulator to metal is due to oxygen vacancy migration near the interface of the electrode and TMO[22]. Therefore, our investigations were focused on this region. The representation and the resulting density of states for each layers of $HfO_2$ slab in contact with the Ta electrodes are shown in Fig. 4. From this, it is noticed that the interface layers of $HfO_2$ (first and second layers) slab have metallic properties and that the lower layers have insulating properties. The interface layers are directly in contact with the metal electrodes and therefore have the greatest tendency to hybridize with the electronic states of the metal electrodes and consequently altering its insulating properties to metallic. With or without the presence of oxygen vacancy, these layers are metallic. These changes in the electronic properties of the TMO through the presence of rowed oxygen vacancy with charge carrier trapping were also observed for the CoO and Ta/CoO systems[34]. These two materials, CoO and $HfO_2$, were known TMO materials used in $RRAM^{TM}$. Therefore, it is assumed that the switching mechanism through the creation of a conduction path is by formation of rowed oxygen vacancy via oxygen vacancy migration from the interface to the lower layers. This switching mechanism is represented in Fig. 5. In the metallic system (low resistance state) of $RRAM^{TM}$ (Fig. 5a), the presence of rowed oxygen vacancy creates a conduction path between interface and lower layers. Application of a sufficient amount of reset pulse voltage moves the oxygen atom to migrate from the interface layers to the lower layers, thus oxygen vacancy migration from the lower layers to the interface layers which switches the system to an insulating system. In the insulating system (high resistance state) of $RRAM^{TM}$ (Fig. 5b), oxygen atom interruption along the conduction path between interface and lower layers makes the system insulating. From this insulating system, an oxygen atom migrates from the lower layers to the interface layers through an application of a sufficient amount of set pulse voltage, which therefore makes the oxygen vacancy migrate from the interface layers to the lower layers thus converting the system back to metallic[7].

Experimental investigations on the set and reset pulse voltage were done to the Ta/HfO$_2$/TiN system. Results in table 1 show that a lower reset pulse voltage is needed to switch the system from a low resistance state to a high resistance state whereas a higher set pulse voltage is needed to convert the system from a high resistance state to a low resistance state. These results were compared with the activation energy barriers for oxygen vacancy migration between the TMOs electrode interface layers and the lower layer region with the inclusion of metal electrodes which, as mentioned before, are essential parts of the switching mechanism. Climbing Nudge Elastic Band (CNEB)[35] method was used to determine the most effective path for oxygen vacancy migration and the corresponding activation energy barrier for oxygen vacancy migration for positions near the electrode-bulk TMO interface of the Ta/HfO$_2$ system. The corresponding activation energy barriers are shown in table 2. Generally, we see that the barrier for conversion from a metallic system (low resistance state) to an insulating system (high resistance state) is less than the barrier for converting it *vice versa*. This is attributed to the fact that TMO is more stably occurring as an insulator than a material with metallic property and therefore easier to switch the system from a metal to an insulator than the other way around. These results were consistent with the results obtained from experimental analysis wherein reset pulse voltage that converts the system from metal to insulator is lower than set pulse voltage which converts the system from insulator to metal, and therefore supports the previously proposed mechanism of switching involving oxygen vacancy migration as the main component of the switching.

**CONCLUSIONS**

This paper has therefore shown a clear picture on the mechanism of the electronic property transition of the TMO which is the main feature of this metal-insulator-metal system of RRAM that makes it one of the most important components of the future non-volatile memory devices. It is clearly shown that this transition is mainly attributed to the creation of an oxygen vacancy row with charge carrier trapping and that this transition mostly occurs near the electrode-bulk interface regions triggered by the application of set and reset pulse voltage.

The ideas conveyed in this paper are of great importance in designing RRAM$^{TM}$ devices consisting of other oxides and electrode materials. We hope that this method serves a useful tool for the development of RRAM$^{TM}$ technology.

**ACKNOWLEDGEMENT**

This work is supported by the New Energy and Industrial Technology Development

Organization's (NEDO) through the project entitled, 'Research and development of transition-metal oxides and its nanofabrication processes for ultra-high-density non-volatile memory'. Some of the calculations presented here are performed using the computer facilities of the Institute of Solid State Physics (ISSP) Super Computer Center (University of Tokyo), the Yukawa Institute (Kyoto University), The Cybermedia Center (Osaka University), and the Japan Atomic Energy Research Institute (ITBL, JAERI). The authors also acknowledge Dr. Wilson Agerico Diño, Dr. Hiroshi Nakanishi and Dr. Sumio Terakawa for fruitful discussions.

## REFERENCES

1. W. Park, J. W. Park, D. Y. Young, and J. K. Lee, J. Vac. Sci. Technol. A **23,** 1309 (2005).

2. S. Seo, M. J. Lee, D. H. Seo, S. K. Choi, D. S. Suh, Y. S. Joung, I. K. Yoo, I. S. Byun, I. R. Hwang, S. H. Kim, and B. H. Park, Appl. Phys. Lett. **86,** 093509 (2005).

3. S. Seo, M. J. Lee, D. H. Seo, E. J. Jeoung, D. S. Suh, Y. S. Joung, I. K. Yoo, I. R. Hwang, S. H. Kim, I. S. Byun, J. S. Kim, J. S. Choi, and B. H.Park, Appl. Phys. Lett. **85,** 5655 (2005).

4. A. Beck, J. G. Bednorz, Ch. Gerber , C. Rossel, D. Widmer, Appl. Phys. Lett. **77,** 139 (2000).

5. S. Q. Liu, N. J. Wu, A. Ignatiev, Appl. Phys. Lett. **76,** 2749 (2000).

6. R. Waser, M. Aono, Nature Mater. **6,** 883-840 (2007).

7. A. Sawa, Materials Today, **11,** 28 (2008).

8. K. Kinoshita, T. Tamura, M. Aoki, Y. Sugiyama, H. Tanaka, Jpn. J. Appl. Phys. 45, L991 (2006).

9. M. N. Kozicki, C. Gopalan, M. Balakrishnan, and M.A. Mitkova, IEEE Trans. Nanotechnol. **5,** 535-544 (2006).

10. C. Schindler, S.C.P.Thermadan, R. Waser, and M.N. Kozicki, IEEE Trans. Electron Devices **54,** 2762-2768 (2007).

11. T. Sakamoto, K. Lister, N. Banno, T. Hasegawa, K. Terabe and M. Aono Appl. Phys. Lett. **91,** 092110 (2007).

12. M. Houssa, M. Tuominen, M. Naili, V. Afanas'ev, A. Stesmans, S. Haukka, M. M. Heyns, J.Appl. Phys. **87,** 8615 (2000).

13. M. J. Rozenberg, I. H. Inoue, M. J. Sánchez, Phy. Rev. Lett. **92,** 178302 (2004).

14. Y. C. Yang, F. Pan, F. Zeng, M. Liu, J. Appl. Phys. **106,** 123705 (2009).

15. J. Ozeki, H. Itoh, J. Inoue, J. Magn. Magn. Mater. **310,** e644 (2007).

16. N. Sasaki, K. Kita, A. Toriumi, Kentaro Kyuno, Jpn. J. Appl. Phys. **48,** 060202 (2009).

17. H. Kishi, T. Kishi, W. A. Diño, E. Minamitani, H. Akinaga, H. Nakanishi, H. Kasai, J. Comput. Theor. Nanosci. **5,** 1976 (2008).

18. Y. Tamai, H. Shima, H. Muramatsu, H. Akinaga, Y. Hosoi, S. Ohnishi, N. Awaya, Extended Abstracts of the 2008 International Conference on Solid State Devices and Materials, Tsukuba (2008) 1166-1167.

19. A. Okiji, H. Kasai, S. Terakawa, J. Phys. Soc. Jpn. **44,** 1275 (1978).

20. A. Okiji and H. Kasai, Surf. Sci. **86,** 529 (1979).

21. M. David, R. Muhida, T. Roman, H. Nakanishi, W. Dino, H. Kasai, F. Takano, H. Shima, H. Akinaga, Vacuum **83,** 599 (2009).

22. G. Kresse, J. Furthmüller, Computer code VASP (Vienna, Austria,1999)

23. G. Kresse, J. Furthmüller, Comput. Mater. Sci. **6,** 15 (1996).

24. G. Kresse, J. Hafner, Phys. Rev. B **47,** 558 (1993).

25. G. Kresse, J. Furthmüller, Phys. Rev. B **54,** 11169 (1996).

26. J. P. Perdrew, J. A. Chevary, S. H. Vosko, K. A. Jackson, M. R. Pederson, D. J. Singh, C. Fiolhais, Phys. Rev. B **46,** 6671 (1992).

27. P.E. Blõchl, Phys. Rev. B **50,** 17953 (1994).

28. G. Kresse, D. Joubert, Phys. Rev. B **59,** 1758 (199).

29. H. J. Monkhorst, J. D. Pack, Phys. Rev. B **13,** 5188 (1976).

30. S. L. Dudarev, G. A. Botton, S. Y. Savrasov, C. J. Humphreys, A.P. Sutton, Phys. Rev. B **57,** 1505 (1998).

31. U. D. Wdowika, D. Legut, Journal of Physics and Chemistry of Solids **69,** 1698 (2008).

32. V. V. Afanas'ev, A. Stesmans, F. Chen, X. Shi, & S. A. Campbell, Appl. Phys. Lett. **81,** 1053 (2002).

33. H. Kasai, T. Kakuda, A. Okiji, Surf. Sci. 363 (1996) 428.

34. H. Kishi, A. A. A. Sarhan, M. Sakaue, S. M. Aspera, M. Y. David, H. Nakanishi, H. Kasai, Y. Tamai, S. Ohnishi, N. Awaya, Jpn. J. Appl. Phys. **50**, 071101 (2011).

35. G. Henkelman, B.P. Uberuaga, and H. Jonsson, Chem. Phys. **113,** 9901 (2000).

**TABLES**

Table 1. Measured voltages for electroforming, and set and reset transition.

|  | Ta/HfO$_2$/Pt |
|---|---|
| TMO Thickness | 5 nm |
| Forming Voltage | 3.3 V |
| Set TransitionVoltage | 0.8 V |
| Reset Transition Voltage | - 0.5 V |

Table 2. Activation barrier for oxygen vacancy migration

| State | HfO$_2$ |
|---|---|
| High resistance to low resistance | 1.3eV |
| Low resistance to high resistance | 0.2eV |

**FIGURE CAPTION**

**Fig. 1**. Supercell used to model bulk $HfO_2$. a. Bulk $HfO_2$, and b. bulk $HfO_2$ with oxygen vacancy row. Dark and gray circles represents O and Hf atoms, respectively.

**Fig. 2.** Electronic property analysis on effect of oxygen vacancy row and charge carrier trapping in $HfO_2$. a-c. Density of States (DOS) vs. Energy relative to the Fermi level, ($E_F$) for: (a.) bulk $HfO_2$, (b.) bulk $HfO_2$ with rowed oxygen vacancy, and (c.) bulk $HfO_2$ with rowed oxygen vacancy and charge carrier traping. d. Band structure near the Fermi level of bulk $HfO_2$ with oxygen vacancy rows and charge carrier trapping. Band overlapping the Fermi level are marked in red. e. Charge density distribution along the band overlapping the Fermi level for a bulk $HfO_2$ with oxygen vacancy row and charge carrier trapping. Moss green, red, and white spheres represents Hf atoms, O atoms, and O atom vacancy, respectively. The colors indicate electron densities of until 0.002 electrons $(Å^3)^{-1}$.

**Fig. 3.** Band structure of single point defect in $HfO_2$. Band structure near the Fermi level of a bulk $HfO_2$ with a single point defect, as a reference to the occurrence of disrupted oxygen vacancy row through oxygen vacancy migration. *Inset:* Illustration of the position of the atoms

and the vacancy of a bulk $HfO_2$. Moss green, red, and white spheres represents Hf atoms, O atoms, and O atom vacancy, respectively.

**Fig 4.** Heterostructure $Ta/HfO_2$ systems and Layer-dependent Local density of states (LDOS) of $HfO_2$ layers. Layer numbers of the $HfO_2$ slabs are shown on the left side. Black, red, and moss green spheres represents, Ta, O, and Hf atoms, respectively. The LDOS of each corresponding layer are shown on the right side. The energies given are relative to the Fermi level.

**Fig. 5.** Schematic illustration of the switching mechanism near the electrode-TMO interface layer. a. Low resistance state of the RRAM$^{TM}$ system. b. High resistance state of the RRAM$^{TM}$ system. Dark gray and dashed line white circles represent O atom and oxygen vacancies, respectively. The center white region represents the conduction path, the light gray region represents the interface TMO layer that is metallic and the dark grey region represents the inner layers of the TMO.

**FIGURES**

Figure 1

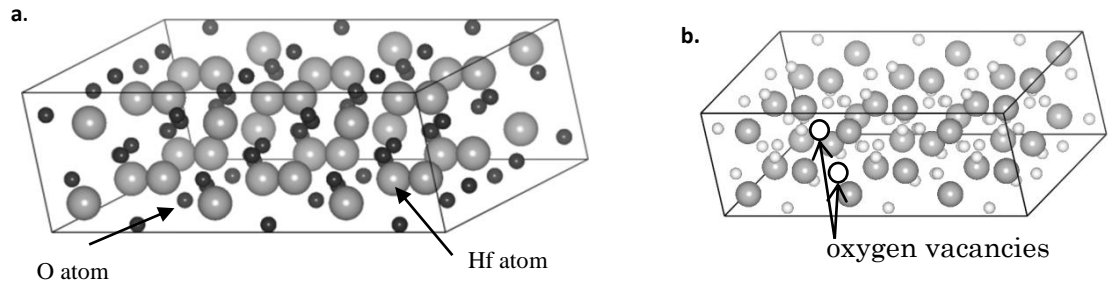

Figure 2

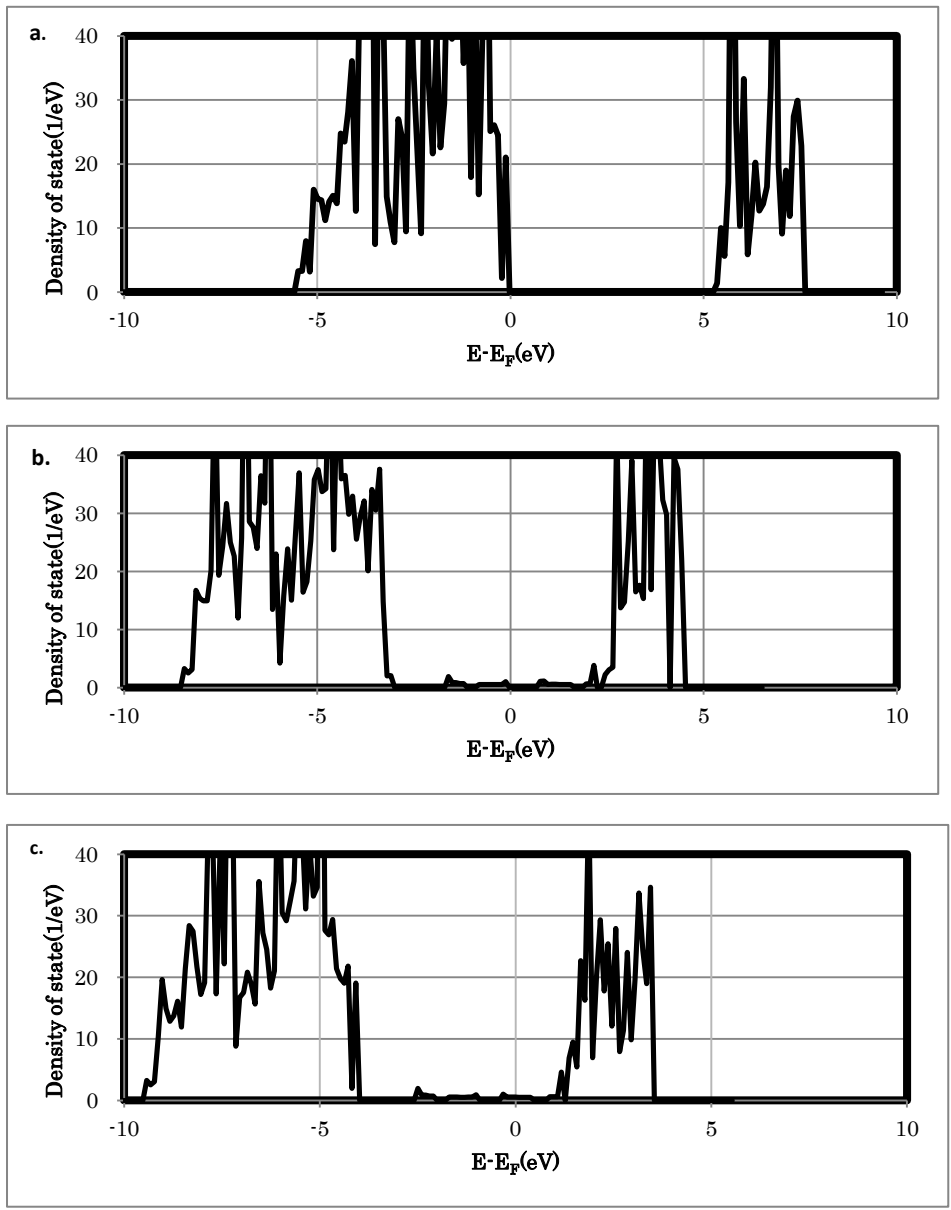

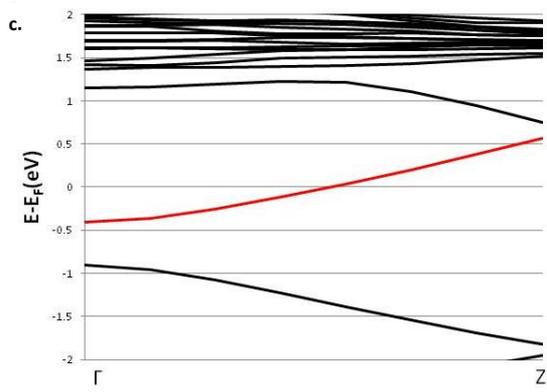

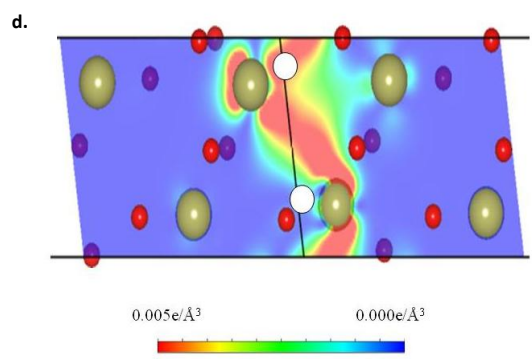

Figure 3

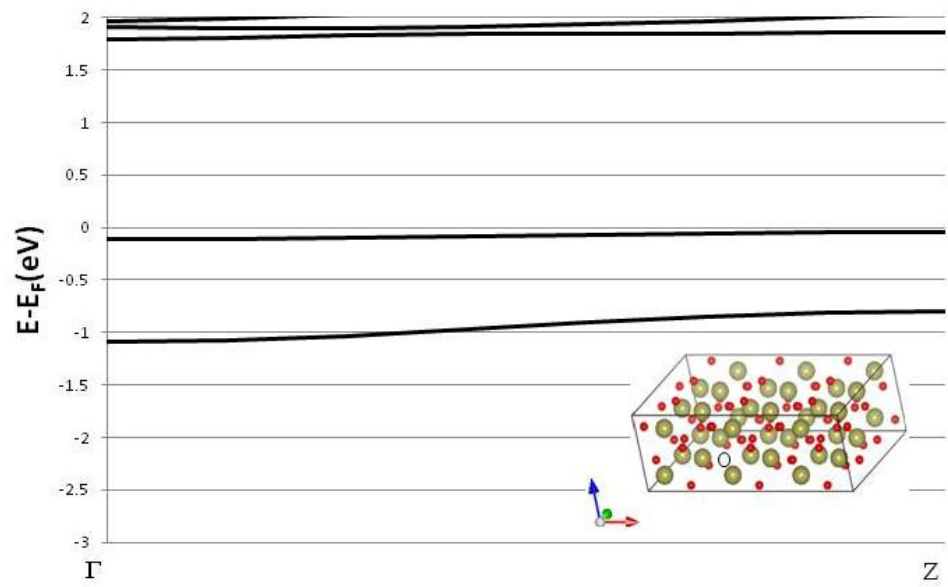

Figure 4

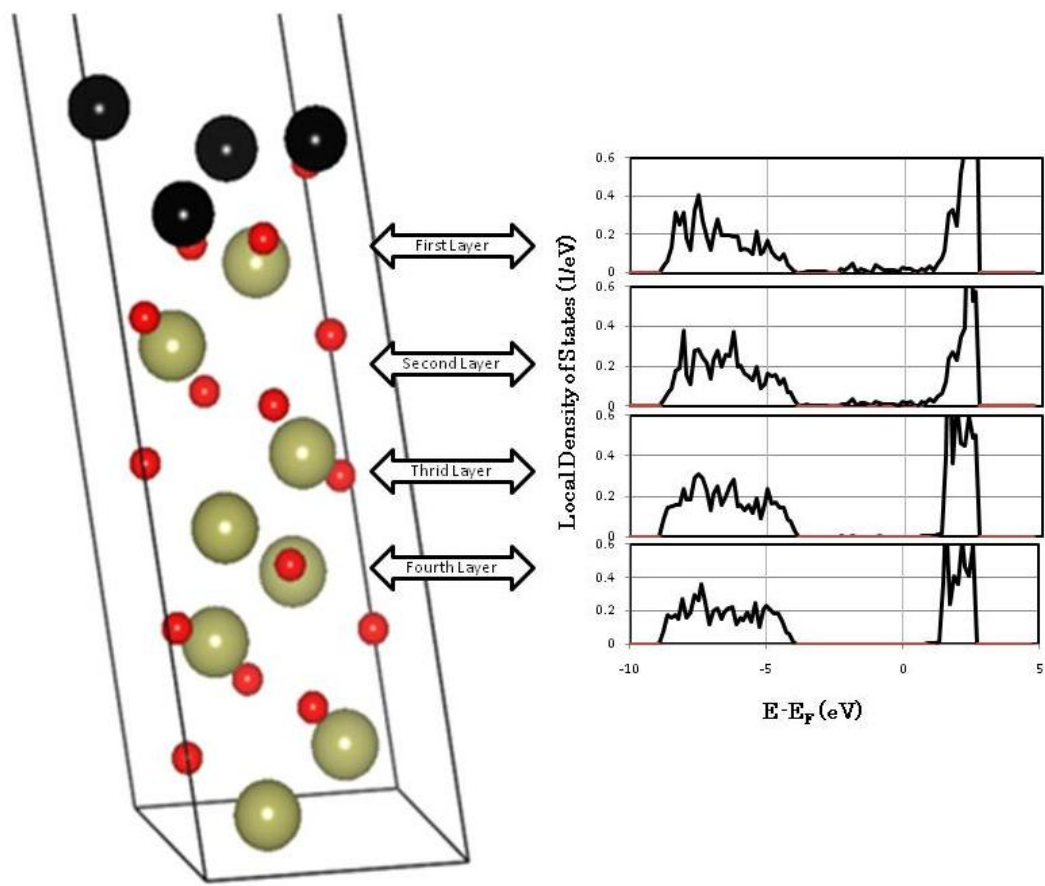

Figure 5

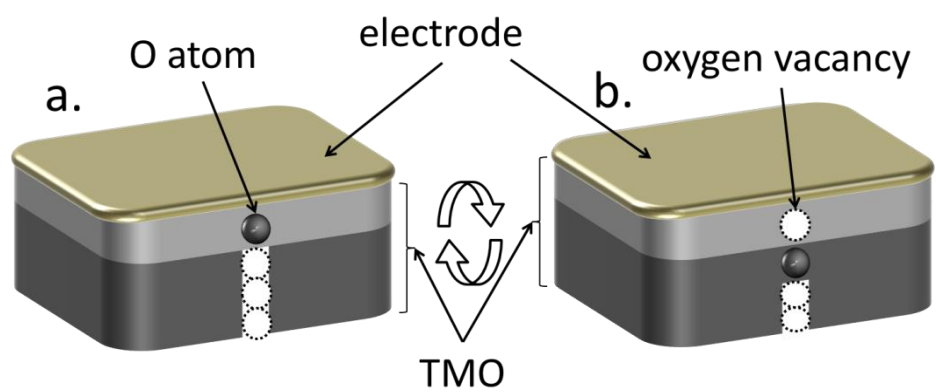